# Study of plasma heating in ohmically and auxiliary heated regimes in spherical tokamak Globus-M.

N.V.Sakharov[1], B.B.Ayushin[1], A.G.Barsukov[2], F.V.Chernyshev[1], V.K.Gusev[1], V.M.Leonov[2], R.G.Levin[1], V.B.Minaev[1], A.B. Mineev[3], M.I. Mironov[1], M.I.Patrov[1], Yu.V. Petrov[1], G.N. Tilinin[2], S.Yu. Tolstyakov[1].

[1]*A.F.Ioffe Physico-Technical Institute, St.Petersburg, Russia*
[2] *Nuclear Fusion Institute, RRC "Kurchatov Institute", Moscow, Russia*
[3] *D.V. Efremov Institute of Electrophysical Apparatus, St. Petersburg, Russia*

## INTRODUCTION

This paper describes the basic features of the plasma heating in spherical tokamak Globus-M. The experiments were performed in the following range of plasma parameters: the toroidal magnetic field 0.4 T near the magnetic axis, the plasma current 0.15-0.25 MA, the plasma average density (1-7) $10^{19}$ $m^{-3}$, the plasma minor radius 0.2-0.24 m, the plasma major radius 0.32-0.36 m, the vertical elongation 1.3-1.8. The neutral beam injection (NBI) was used as the basic method for the plasma auxiliary heating. In the experiments the injector could generate a hydrogen or deuterium beam with the energy of 30 keV and the neutral beam power of 0.6 MW. The beam was co-injected relatively the plasma current direction in the tokamak midplane. The beam axis was aimed into the inner plasma region at the radius R = 0.3 m. The experiments were carried out after a careful beam focusing and outgasing of the the vacuum vessel inlet port. We employed the glow discharge cleaning and the boronization technique [1] for the vacuum vessel conditioning. About 50% of the plasma faced surface was protected with graphite tiles. After the boronization the plasma radiation losses did not exceed 10-15% of the ohmic power. In the described experiments the most interesting effects were observed in the ion temperature behavior in ohmically heated plasmas as well as during neutral beam injection. For this reason the present paper is mostly concentrated on the features of the ion energy balance. The ion temperature in the plasma core was measured by means of the 12-channels neutral particle analyzer (NPA) ACORD-12. The NPA provided the simultaneous measurements of deuterium and hydrogen energy spectra and the percentage of both isotopes. The electron temperature profile was measured with the laser Thomson scattering diagnostic in five spatial points positioned along the plasma major radius. The electron density was monitored by 1 mm interferometer along three vertical chords. The estimates of electron temperature in the plasma core were derived also from the slope of SXR spectra.

**OH Regime**

Some characteristic features in the ion temperature behavior were described in [2]. It was found that the ion temperature in the plasma core appeared to be higher than the values predicted by the Artsimovich scaling law [3] $T_i \sim (nI_pB_tR^2)^{1/3}/A_i^{0.5}$ well describing the ion heating in the case of the neoclassical plateau regime in the conventional tokamak. The experiments revealed much stronger, almost linear dependence of $T_i$ upon the plasma current. This dependence was well confirmed by 1D transport code using the neoclassical transport coefficients.

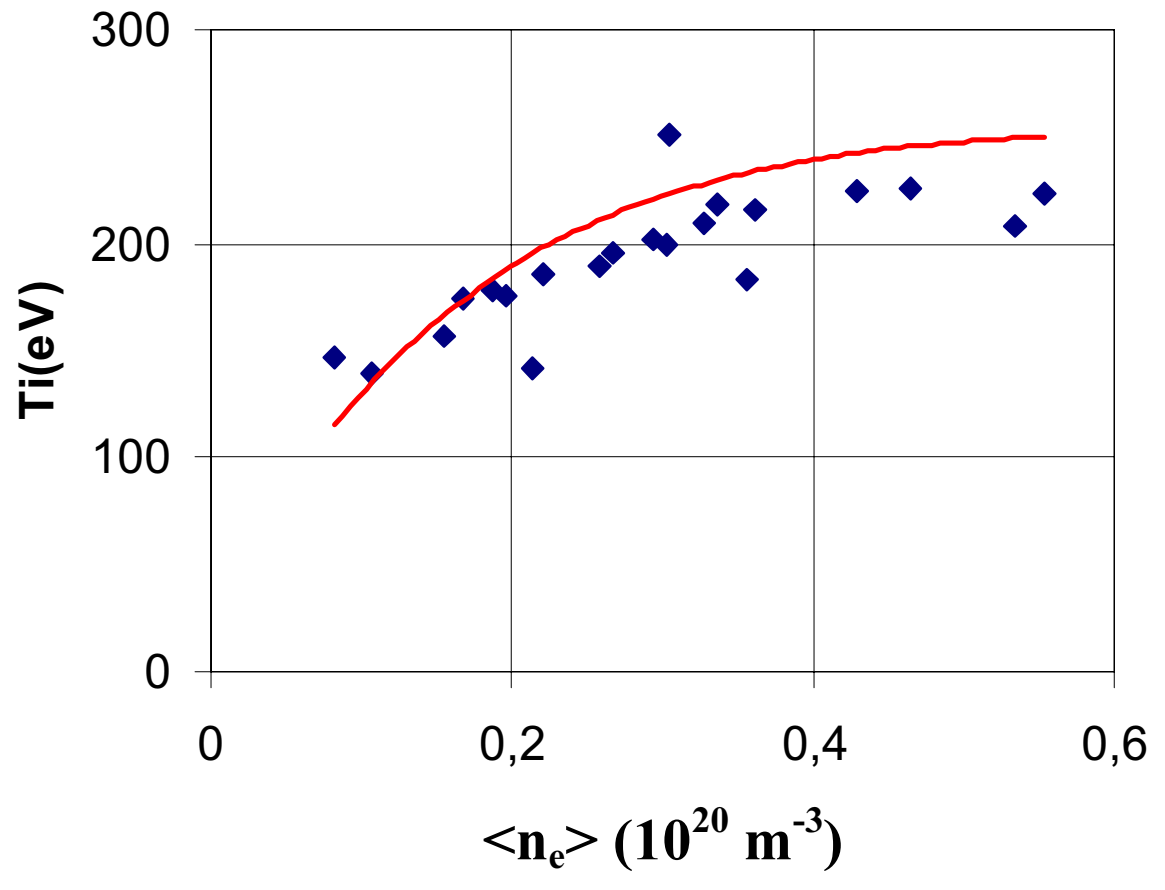

*Fig.1. Ion temperature in the plasma core as a function of average density. $I_p$ = 215 kA. Solid line shows the results of modeling.*

In this work we continued the study of ion temperature dependence upon the plasma parameters. Fig.1 shows the ion temperature in the plasma core as a function of the plasma average density. The density control was provided by the gas (deuterium) puffing from the high magnetic field side. In the experiment the density was scanned in the range of $(0.8 - 5.5)\times10^{19}$ $m^{-3}$. Note that the temperature obtained from the slope of charge-exchange particles energy spectra gives the lower estimate of the central ion temperature especially at a higher density, because these so called "passive" spectra are the integral functions over the analyzer line of sight.

The solid line in Fig.1 demonstrates the result of numerical simulation by means of 1D transport model. In the model we used Hinton-Hazeltine transport coefficient [4]:

$$\chi_{iHH} = K_2 \cdot \frac{r_{Bi}^2 \cdot q^2}{\varepsilon^{3/2}} \cdot \nu_{ii}\ ;\ K_2 = 0.66 \cdot \left\{ \frac{1}{1 + 1.03\sqrt{\nu_{*i}} + 0.31 \cdot \nu_{*i}} + \frac{1.77 \cdot \varepsilon^3 \cdot \nu_{*i}}{1 + 0.74 \cdot \varepsilon^{3/2} \nu_{*i}} \right\}$$

where $r_{Bi}$ – ion gyroradius, $\varepsilon = r/R$, $\nu_{ii}$ – ion-ion collision frequency, $\nu_{*i}$ – ratio of ion-ion collision frequency to banana orbit frequency.

In the model the electron temperature and density radial distributions were approximated by the expressions: $T_e = T_{e0}(1-r^2/a^2)^{\alpha T}$, $n_e = n_{e0}(1-r^2/a^2)^{\alpha n}$, where the central temperature $T_{e0}$ and the coefficients $\alpha_T$ and $\alpha_n$ were chosen to fit the experimentally measured electron temperature profile and density measurements along vertical chords. The electron temperature radial distribution measured in five spatial points along the plasma major radius is shown in Fig.2. We observed a weak dependence of the electron temperature upon the plasma density in the experimental density range shown in Fig.1. In addition in most plasma shots the values of $T_e(0)$ were approximately by factor of 2 larger of the $T_i$ values. Under these conditions the electron-ion heat flux is a weak function of the electron temperature, and the shape of the $T_e$ radial distribution is not so important for the ion energy balance in the plasma center. The model included charge-exchange loses which were determined by the neutral density near the plasma boundary and the neutral particle influx from the plasma edge to the plasma core. In the simulation the edge neutral density $n_0 = 3\times10^{15}$ $m^{-3}$ caused a variation of the ion temperature in the plasma center within 10%. At the same time it made the modeling low sensitive to the boundary values of the plasma density and temperature. The ion convective losses were not taken into account in the modeling. The typical simulated ion temperature radial distribution is shown in Fig.3. Under the above mentioned conditions the simulation using the neoclassical transport coefficients well describes the experimentally measured density dependence of the central ion temperature in all the experimental density range (see solid line in Fig.1). Note that in the case of dominating charge-exchange losses the neoclassical model predicts much stronger density dependence of the ion temperature. From the model we can estimate the ion energy confinement time determined as $\tau_{Ei} = W_i/P_{e\text{-}i}$, where $W_i$ – volume integrated ion thermal energy, $P_{e\text{-}I}$ – volume integrated electron-ion heat power. The $\tau_{Ei}$ value appeared to be a weak function of the plasma density which changed within 6÷7 ms in the investigated density range.

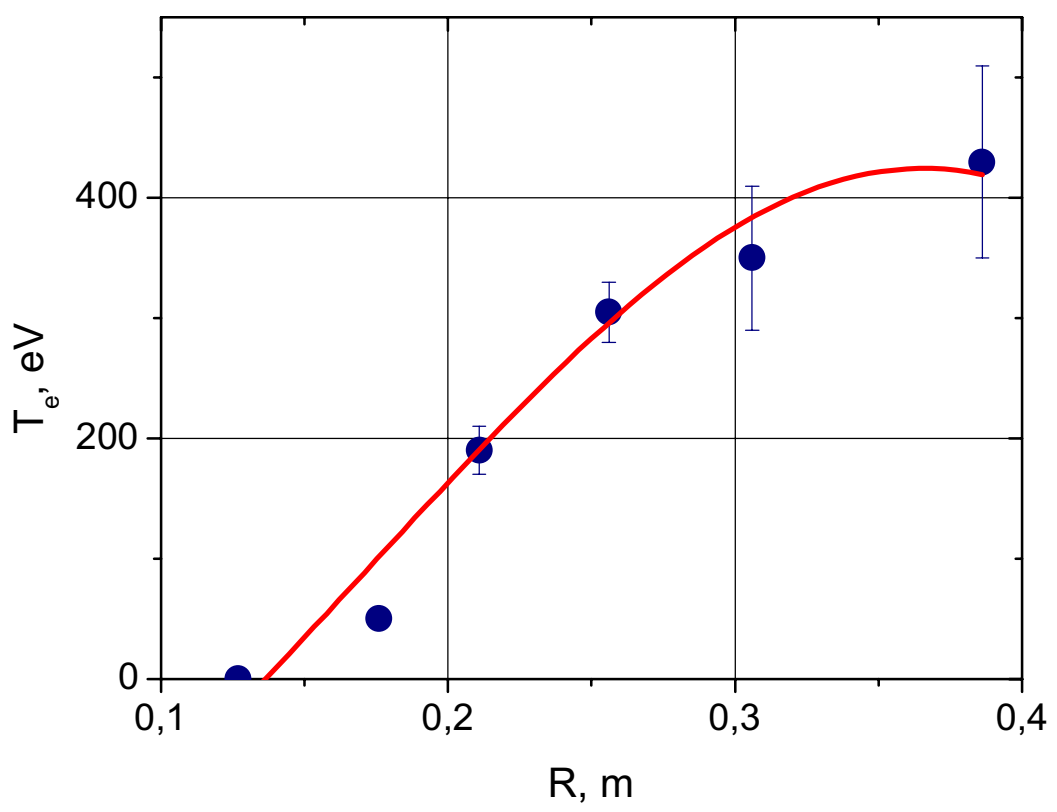

*Fig.2. Electron temperature profile measured by Thomson scattering diagnostic.*

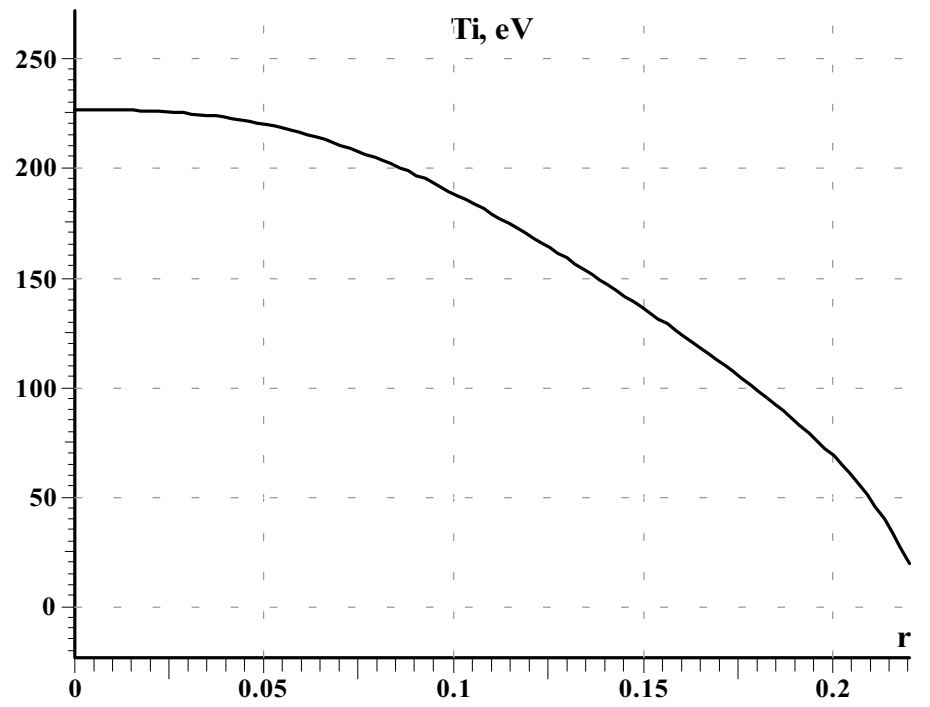

*Fig.3. Calculated ion temperature profile. Minor radius r(m). Plasma average density* $\langle n_e \rangle = 0.34\times10^{20}$ $m^{-3}$

**NB heating**

The NBI technique in Globus-M experiment is described in [5]. In the experiments described in this paper we compared the effect of the plasma heating by means of the hydrogen and deuterium beams injection into deuterium plasmas. This comparative experiment was carried out during one experimental session. In both cases the beam parameters were approximately the same: 0.6 MW of the beam power and 30 keV of the beam energy. For the hydrogen beam it means that the particle velocity was by a factor of $2^{0.5}$ higher than for the deuterium beam. The deuterium temperature time evolution in OH regime and in the regimes with the hydrogen and deuterium beams injection is shown in Fig.4. The corresponding deuterium charge-exchange energy spectra are shown in Fig.5. The ion temperature was determined in the range of energies below 1.5 keV. The experiments were performed in the plasma discharge before the vacuum vessel boronization. The beam of 20 ms duration was injected in the phase of a steady state plasma current of 0.15-0.16 MA. In both cases of the hydrogen and the deuterium beam we observed approximately the same effect of ion heating. The beam injection was accompanied by a factor of 2-2.5 rise of the plasma density and the appearance of sawtooth oscillations. These phenomena can be a possible explanation of the saturation and then the following decrease in the ion temperature during the NBI pulse. At the same time we did not observe any significant variations in the emission of typical impurity lines such as CIII, OIII etc, that indicated the absence of plasma additional contamination with light impurities during NBI. The electron temperature in the plasma bulk estimated from the slope of SXR spectra also did not change significantly in NBI experiments.

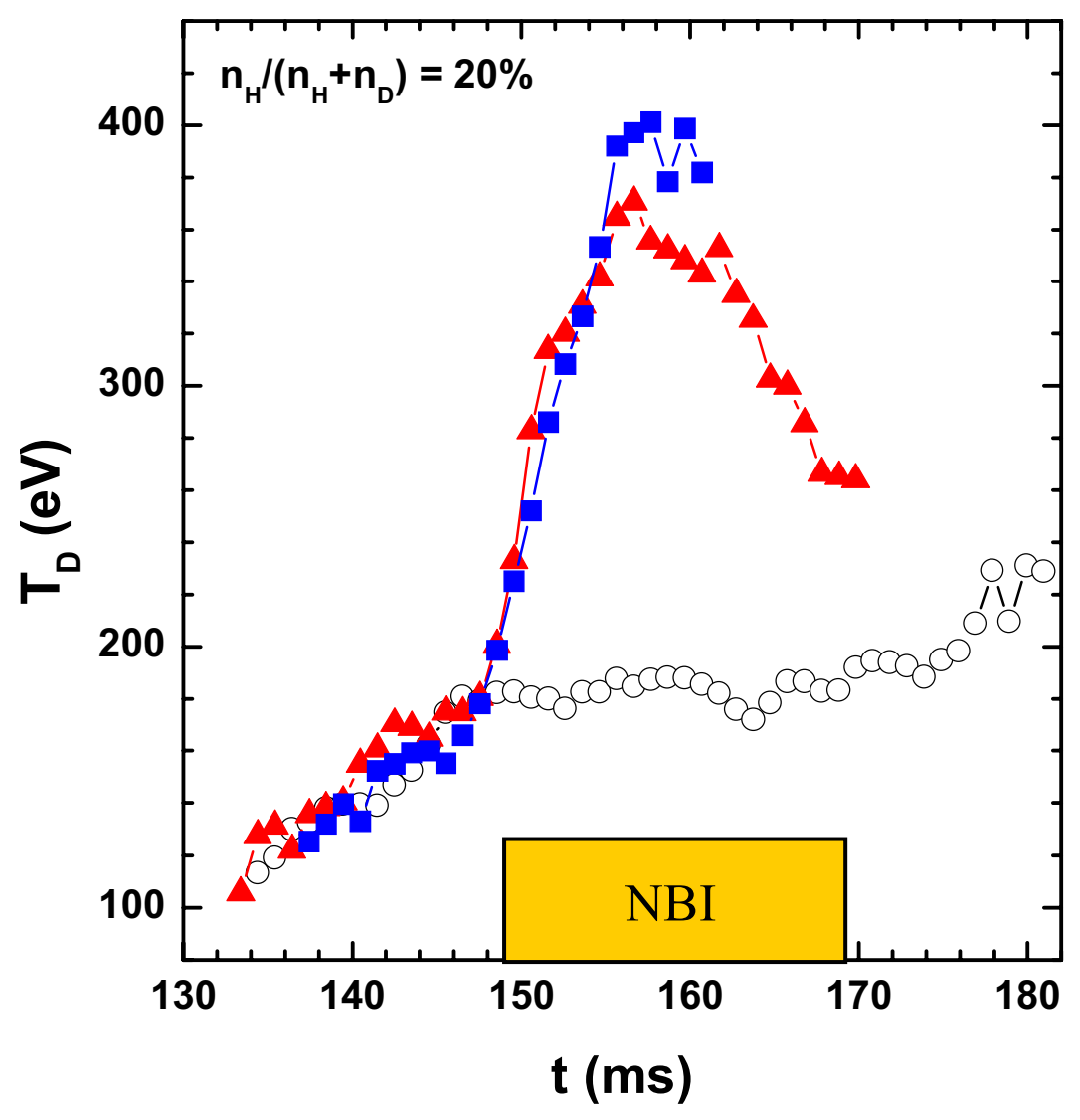

*Fig.4. Time evolution of $T_i$ in plasma center in OH and NBI regimes.*
*$I_p$ = 0.15-0.16 MA, $P_{NBI}$ = 0.6 MW*
*○ – OH, ▲ – H-NBI, ■ – D-NBI*

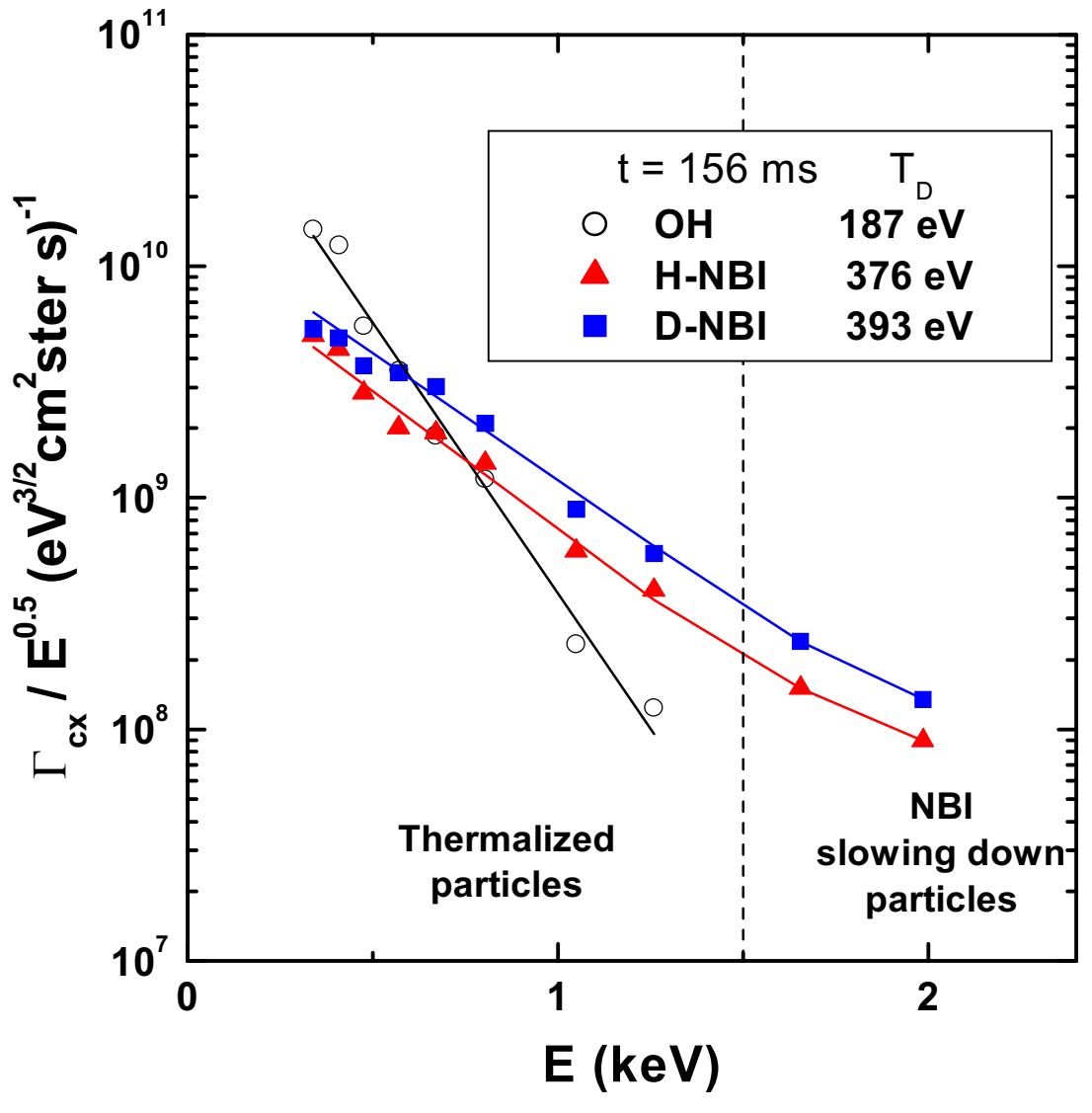

*Fig.5. Deuterium energy spectra in OH and NBI regimes*

**Discussion and Conclusions**

The ion temperature behavior in the plasma bulk in OH experiments is well described by the neoclassical transport coefficients without any additional correction. The absolute values and the $T_i$ parametric dependences differ from the Artsimovich scaling law for the neoclassical plateau regime. The modeling demonstrates small ion collisionality in wide zone in the plasma volume in most Globus-M ohmic experiments: $\nu_{*i} = \nu_{ii}/\nu_{bi} << 1$, where $\nu_{bi}$ – banana orbit frequency. It indicates that the trapped particles dominate in the ion transport losses in spherical tokamak even in a relatively low temperature range.

The effects of NBI are well described by a preliminary ASTRA code simulation [6]. A typical balance of the hydrogen beam power as a function of plasma density is shown in Fig.6. In the experiment shown in Fig.4 the beam was injected into the plasma with relatively low initial density <n> = (1.5-2)×$10^{19}$ $m^{-3}$. In this density range the beam power absorbed by electrons and by ions (Fig.6) is about 10-15% of the total beam power. This power is much smaller of the ohmic heating power and therefore it can’t provide a significant increase in the electron temperature. From the experiment we can roughly estimate the power absorbed by ions as $P_{bm_i} \sim \Delta W_i/\Delta t$, where $\Delta W_i \sim \Delta <nT_i>$ is a variation of the ion thermal energy during the starting NBI period. This estimate for the case in Fig.4 gives the value of $P_{bm_i} \sim$ 60-80 kW that well agrees with ASTRA code simulation. The value of the electron-ion heat power in ohmically heated plasma calculated by means of the transport code $P_{e\text{-}i}$ is about 20-25 kW which is small in comparison with $P_{bm_I}$. It explains a strong ion temperature increase in Fig.4 for the case of the beam injection into a low density plasma. In the case of higher densities the values of $P_{bm_i}$ and $P_{e\text{-}I}$ become comparable, and the experimentally observed increase of $T_i$ is smaller. In Fig.6 the difference between the absorbed beam power $P_{abs}$ and the total power lunched by electrons and ions $P_{bm_ei}$ is described by the beam particle particles velocity charge-exchange losses. A significant role of the charge exchange losses in NB power balance is partly confirmed by the experimentally observed decrease of the ion temperature in plasma shots immediately after the vessel boronization. After the boronization a strong external gas puffing is necessary to sustain the plasma density. Under these conditions the NPA detects a significant rise of neutral fluxes.

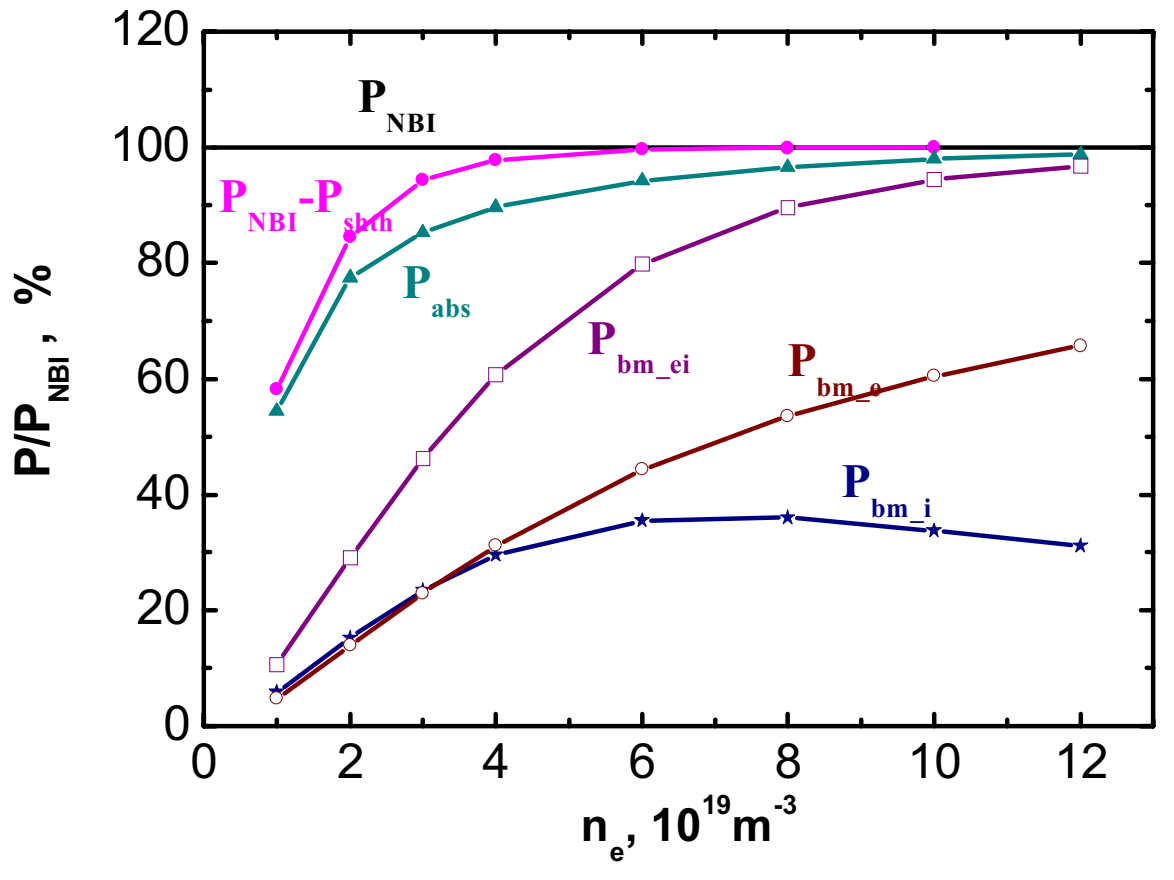

*Fig.6. NBI absorbed power as a function of average plasma density*
*$P_{NBI}$ = 0.6 MW, $E_{NBI}$ = 30 keV*
*$P_{NBI}$ – total input power, $P_{NBI}$ – $P_{shth}$ – total input power without shine-through losses, $P_{abs}$ – total input power without shine-through and first-orbit losses, $P_{bm_e}$ – power absorbed by electrons, $P_{bm_i}$ – power absorbed by ions,*
*$P_{bm_ei}$ – power absorbed by electrons and ions*

**Acknowledgements.**
The work is supported by Russian Academy of Sciences, Russian Ministry of Education and Science, Russian Agency of Atomic Energy and by RFBR grants 02-02-17693, 03-02-17659.

**References.**